\documentclass[aps,twocolumn,amsmath,amssymb,preprintnumbers,superscriptaddress,floatfix]{revtex4-2}

\usepackage[version=4]{mhchem}
\usepackage[utf8]{inputenc}
\usepackage{newtxtext}
\usepackage[upint]{newtxmath}
\usepackage{microtype}
\usepackage{textcomp}
\usepackage{dsfont}
\usepackage{eucal}
\usepackage{siunitx}
\usepackage{soul}

\usepackage{enumerate}
\usepackage{amsfonts}
\usepackage{color}
\usepackage{soul}

\usepackage{todonotes}
\presetkeys%
    {todonotes}%
    {inline}{}

\usepackage{graphicx}

\usepackage[colorlinks,allcolors=blue]{hyperref}
\usepackage[capitalize]{cleveref} % Do we want to capitalize?
\usepackage{cleveref}

\definecolor{DarkBlue}{rgb}{0,0,0.80}
\definecolor{DarkRed}{rgb}{0.80,0,0}
\definecolor{Purple}{rgb}{0.55,0,0.55}
\definecolor{Purple}{rgb}{0,0,0.8}

\let\epsilon\varepsilon

\begin{document}

\title{Barrier and finite size effects on the extension of topological surface-states into magnetic insulators}
\author{Eirik Holm Fyhn}
\affiliation{Center for Quantum Spintronics, Department of Physics, Norwegian \\ University of Science and Technology, NO-7491 Trondheim, Norway}
\author{Hendrik Bentmann}
\affiliation{Center for Quantum Spintronics, Department of Physics, Norwegian \\ University of Science and Technology, NO-7491 Trondheim, Norway}
\author{Jacob Linder}
%\email[Corresponding author: ]{jacob.linder@ntnu.no}
\affiliation{Center for Quantum Spintronics, Department of Physics, Norwegian \\ University of Science and Technology, NO-7491 Trondheim, Norway}\date{\today}
\begin{abstract}
The interplay between magnetic and topological order can give rise to phenomena such as the quantum anomalous Hall effect. The extension of topological surface states into magnetic insulators (MIs) has been proposed as an alternative to using intrinsically magnetic topological insulators (TIs). Here, we theoretically study how this extension of surface states into a magnetic insulator are influenced both by the interface barrier potential separating a topological insulator and a magnetic insulator and by finite size effects in such structures. We find that the the gap in the surface states depends non-monotonically on the barrier strength. A small, but finite, barrier potential turns out to be advantageous as it permits the surface states to penetrate even further into the MI. Moreover, we find that due to finite size effects in thin samples, increasing the spin-splitting in the MI can actually decrease the gap of the surface states, in contrast to the usual expectation that the gap opens as the spin-splitting increases.
\end{abstract}

\maketitle
\section{Introduction}%
\label{sec:introduction}
 Topological insulators (TIs) are materials that are insulating in the bulk and have topologically protected conducting surface states \cite{konig:07.2,Hasan:10.11,Qi:11}. 
Combining the topological properties of TIs with magnetic order can give rise to the quantum anomalous hall effect (QAHE) as a result of broken time-reversal symmetry in the TI \cite{yu2010,Qi:11,Nomura:11,Tokura2019,bernevig2022}, which could have applications in low-dissipation spintronics devices.
Magnetic TIs are also discussed as an avenue to the realization of Majorana quasi-particles with potential relevance for implementations of quantum-computation platforms \cite{Tokura2019}.
The first observation of the QAHE was in 2013, where the effect was achieved by doping a TI with magnetic impurities \cite{Chang2013,Chang2015}.
However, achieving QAH resistance quantization at zero magnetic field has remained limited to temperatures of the order of tens of mK in these systems, which was explained by thermally activated bulk conductance \cite{fijalkowski2021}.
An interesting alternative way to produce the QAHE would be to use an intrinsic magnetic TI, such as the van der Waals magnet \ce{MnBi2Te4} \cite{otrokov:19}. Yet, while signatures of the QAHE were reported in thin exfoliated flakes of odd-layer thickness \cite{deng_science_20}, the antiferromagnetic order in \ce{MnBi2Te4}, in general, does not provide ideal prerequisites for a robust QAHE.

Another approach to break time-reversal symmetry at the surface of a TI is to proximize the TI to a magnetic insulator (MI) \cite{Qi:11}. This should induce an exchange gap in the Dirac cone of the topological surface state. Various TI-MI heterostructures have been investigated experimentally \cite{liu2023}, but the observation of magnetic topological behavior and of an exchange gap in the surface state has remained challenging. More recently, it has been proposed that the topological surface state can be extended into the MI in van der Waals heterostructures with weak potential modulation at the interface \cite{otrokov2017highly,otrokov:17_magnetic}.
That is, instead of inducing a weak magnetization at the surface of the TI, interface effects give rise to a relocation of the surface state from the TI into the MI. This ferromagnetic extension requires similar atomic structures for the TI and MI. Compared to  a mere proximity effect, such relocation of the surface state into the MI would drastically enhance the exchange gap \cite{otrokov2017highly,otrokov:17_magnetic}.
A ferromagnetic extension was experimentally achieved in \ce{Bi2Te3}/\ce{MnBi2Te4} heterostructures \cite{kagerer_prr_23}, where it was possible to observe an exchange gap in the surface state  up to the critical temperature of 15~K.

Here we theoretically study how the topological surface states in TI/MI heterostructures are affected by the barrier potential at the TI/MI interface, the magnitude of the spin-splitting, and finite size effects due to the finite thickness of the materials. 
We find that the energy gap of the surface states has a non-monotonic dependence on the barrier strength. Below a critical value, the barrier potential actually turns out to be advantageous as it permits the surface states to penetrate further into the MI compared to an ideal, barrier-free interface. In addition, we find that due to finite size effects in thin samples, an increase in the spin-splitting in the MI can diminish the gap of the surface states at certain thicknesses. This stands in contrast to the usual expectation that the gap opens as the spin-splitting increases. These results might prove useful for optimizing experimental efforts aiming to achieve extension of topological surface states into a time-reversal symmetry-breaking magnetic environment. 

\section{Model}%
\label{sec:model}
To model the TI/MI bilayer we adopt an effective 4-band low-energy model~\cite{zhang2009} for momenta near the $\Gamma$ point,
\begin{equation}
  H = \varepsilon_k + V +
  \begin{pmatrix}
    M_k \tau_z + A_k \tau_x + h & B_k \tau_x \\
    B_k^* \tau_x & M_k\tau_z - A_k\tau_x - h
  \end{pmatrix},
  \label{eq:hamiltonian}
\end{equation}
where the basis is a set of four orbitals,
\begin{equation}
  \Psi = \begin{pmatrix}
    \lvert P1_z^+, \uparrow \rangle & 
    \lvert P2_z^-, \uparrow \rangle & 
    \lvert P1_z^+, \downarrow \rangle & 
    \lvert P2_z^-, \downarrow \rangle 
  \end{pmatrix}^T,
\end{equation}
For simplicity, we have assumed the same $g$-factor for both orbitals in \cref{eq:hamiltonian}.

The parameters $\varepsilon_k$, $M_k$, $A_k$, and $B_k$ are functions of momentum parallel to the interface, $k_\parallel$ and momentum orthogonal to the interface, $k_z$.
As the translational symmetry is broken in the direction orthogonal to the interface, we perform a Peierls substitution, $k_z \to -i\partial_z$.
With this,
\begin{subequations}
  \begin{align}
    \varepsilon_k &= C_0(z)  + \frac 1 2 \left\{C_1(z),\, (-i\partial_z)^2\right\} + C_2(z) k_\parallel^2,\\
    M_k &= M_0(z)  + \frac 1 2 \left\{M_1(z),\, (-i\partial_z)^2\right\} + M_2(z) k_\parallel^2,\\
    A_k &= \frac 1 2 \left\{A_1(z),\, -i\partial_k\right\}, \\
    B_k &= B_1(z)(k_x + ik_y),
  \end{align}
\end{subequations}
where $\{C_j, M_j, A_1, B_1\}$ are material specific parameters and $j\in\{0,1,2\}$.
The anticommutators are added to ensure that the Hamiltonian is Hermitian.
We let the interface be located at $z = 0$ and to allow for a smooth numerical solution we use a sigmoid function,
\begin{equation}
  \theta(z) = \frac{1}{1+\mathrm{e}^{-x/\SI{0.14}{\angstrom}}},
\end{equation}
to define the transition between the TI and the MI that extends over a length of around $\SI{1}{\angstrom}$.
That is, 
\begin{equation}
  M_0(z) = M_0^L (1-\theta(z)) + M_0^R\theta(z),
\end{equation}
and similarly for all the other parameters, where $\{C_j^L, M_j^L, A_1^L, B_1^L\}$ is the set of material parameters for the TI and $\{C_j^R, M_j^R, A_1^R, B_1^R\}$ is the set of material parameters for the MI.
Additionally, \cref{eq:hamiltonian} includes the Zeeman exchange energy
\begin{equation}
  h(z) = h_0 \theta(z),
\end{equation}
and a potential barrier
\begin{equation}
  V(z) = V_0 (\partial_z\theta)(z).
\end{equation}
Note that
\begin{equation}
  \int_{-\infty}^\infty \mathrm{d}z\, V(z) = V_0
\end{equation}
by construction.

For the TI we choose parameters that correspond to \ce{Bi2Se3}, meaning that
$C_0^L = \SI{-6.8e-3}{\electronvolt}$, $C_1^L = \SI{1.3}{\electronvolt\angstrom\squared}$, $C_2^L = \SI{19.6}{\electronvolt\angstrom\squared}$, $M_0^L = \SI{0.28}{\electronvolt}$, $M_1^L = \SI{-10}{\electronvolt\angstrom\squared}$, $M_2^L = \SI{-56.6}{\electronvolt\angstrom\squared}$, $A_1^L = \SI{2.2}{\electronvolt\angstrom}$, and $B_1^L = \SI{4.1}{\electronvolt\angstrom}$.
We vary the parameters of the MI, but set $C_0^R = \SI{-4.8e-3}{\electronvolt}$, $C_1^R = \SI{2.7232}{\electronvolt\angstrom\squared}$, $C_2^R = \SI{17}{\electronvolt\angstrom\squared}$, $M_1^R = \SI{-11.9048}{\electronvolt\angstrom\squared}$, $M_2^R = \SI{-9.4048}{\electronvolt\angstrom\squared}$, $A_1^R = \SI{2.7023}{\electronvolt\angstrom}$, and $B_1^R = \SI{3.1964}{\electronvolt\angstrom}$ fixed.
These values correspond to \ce{MnBi2Te4} in the antiferromagnetic state~\cite{zhang_prl_19}.
We also set $k_y = 0$, such that $k_\parallel = k_x$.
We denote by $L_{TI}$ and $L_{MI}$ the lengths of the TI and MI respectively.

To numerically determine eigenstates and eigenvalues of the Hamiltonian, we discretize the wave functions in space and approximate derivatives by finite differences.
In order to ensure sufficient resolution at the interface we also perform a substitution $\zeta = f(z)$ and discretize $\zeta$ such that a uniform $\zeta$-grid correspond to a high density of discretization points near $z = 0$.

\section{Magnetic extension of surface states}
\label{sec:extension}
\begin{figure}[htpb]
  \centering
  \includegraphics[width=0.48\textwidth]{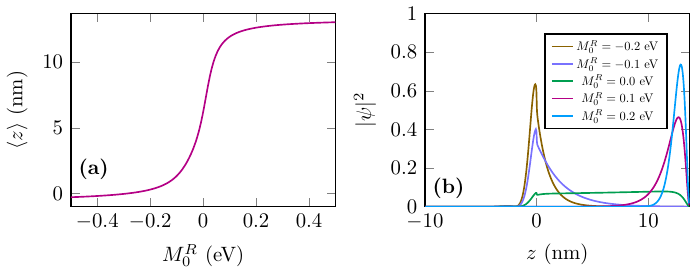}
  \caption{\textbf{(a)} Penetration depth, $\langle z \rangle$, associated with the surface states as a function of $M_0^R$ with $h_0 = \SI{0.1}{\electronvolt}$, $V_0 = 0$, $L_{TI} = \SI{10}{\nano\meter}$ and $L_{MI} = \SI{13.7}{\nano\meter}$. \textbf{(b)} The square amplitude of the interfacial surface state at a selection of $M_0^R$ values and the same parameters as in \textbf{(a)}.}
  \label{fig:penetrationvM}
\end{figure}

First, we consider how the magnetic extension of surface states varies as a function of $M^R_0$.
\Cref{fig:penetrationvM} shows the penetration depth,
\begin{equation}
  \langle z \rangle = \int_{-L_{TI}}^{L_{MI}} \mathrm{d}z\, z\lvert \psi(z) \rvert^2,
\end{equation}
where $\psi$ is the wave function corresponding to a surface state.
We set $k_x = 0$ when evaluating wave functions of surface states.
At $h_0=0$ and $k_x=0$ the Hamiltonian is block-diagonal and the two blocks are related through time-reversal.
Therefore, the eigenfunctions come in pairs that are related through time reversal.
For this reason, there are two surface states at both surfaces of the TI.
The two surface states are no longer Kramer pairs when $h\neq 0$, but we find that they have similar penetration depth and spatial distribution for the parameters in \cref{fig:penetrationvM}.
Therefore, although \cref{fig:penetrationvM} only shows the penetration depth of one of the surface states, the results are similar for the other surface state.

\Cref{fig:penetrationvM} shows a sudden jump in $\langle z \rangle$ as $M_0^R$ crosses $0$.
This is reasonable because $M_0^R = 0$ marks the transition from topologically trivial ($M_0^R/M_2^R > 0$) to topologically nontrivial ($M_0^R/M_2^R < 0$).
The MI is topologically trivial when $M_0^R < 0$, and the surface state is therefore located at the interface between the TI and the MI at $z = 0$.
As $\lvert M_0^R\rvert$ is reduced, the bulk gap is reduced and therefore energy difference between the surface states and the bulk states is reduced.
This leads to the surface states becoming more delocalized until the state is fully delocalized inside the MI when $M_0^R = 0$.
As $M_0^R$ is increased further, the surface state again becomes more localized, but now it centers around the interface between the MI and vacuum because the MI is also topologically nontrivial \cite{Qi:11}

\begin{figure}[htpb]
  \centering
  \includegraphics[width=0.45\textwidth]{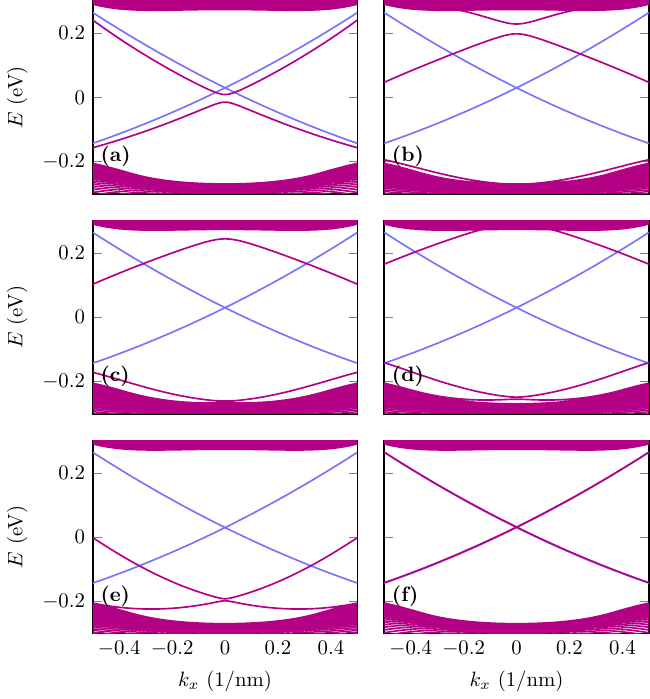}
  \caption{Energy spectra for TI/MI heterostructures with $L_{TI} = L_{MI} = \SI{50}{\nano\meter}$, $M_0^R = \SI{-0.5}{\electronvolt}$, $h_0 = \SI{50}{\milli\electronvolt}$, and \textbf{(a)} $V_0=\SI{0}{\electronvolt\angstrom}$, \textbf{(b)} $V_0=\SI{0.3}{\electronvolt\angstrom}$, \textbf{(c)} $V_0=\SI{0.35}{\electronvolt\angstrom}$, \textbf{(d)} $V_0=\SI{0.4}{\electronvolt\angstrom}$, \textbf{(e)} $V_0=\SI{0.6}{\electronvolt\angstrom}$, \textbf{(f)} $V_0=\SI{10}{\electronvolt\angstrom}$. The blue lines correspond to the surface states at $z = -L_{TI}$.}
  \label{fig:spectravV}
\end{figure}

\begin{figure}[htpb]
  \centering
  \includegraphics[width=0.45\textwidth]{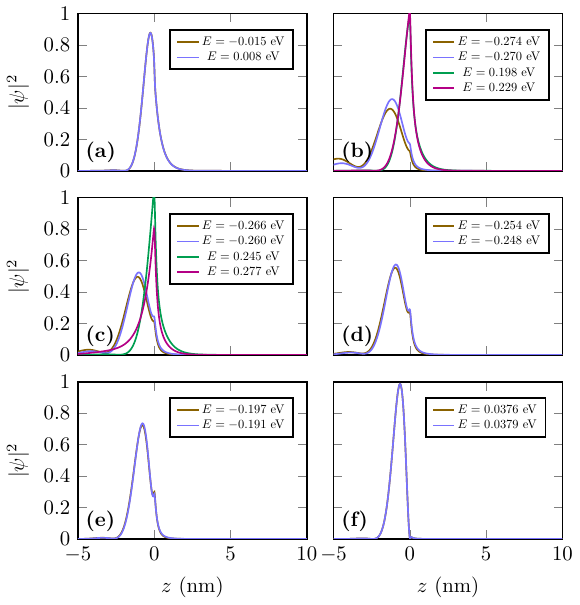}
  \caption{Interfacial surface states in TI/MI heterostructures with $L_{TI} = L_{MI} = \SI{50}{\nano\meter}$, $M_0^R = \SI{-0.5}{\electronvolt}$, $h_0 = \SI{50}{\milli\electronvolt}$, and \textbf{(a)} $V_0=\SI{0}{\electronvolt\angstrom}$, \textbf{(b)} $V_0=\SI{0.3}{\electronvolt\angstrom}$, \textbf{(c)} $V_0=\SI{0.35}{\electronvolt\angstrom}$, \textbf{(d)} $V_0=\SI{0.4}{\electronvolt\angstrom}$, \textbf{(e)} $V_0=\SI{0.6}{\electronvolt\angstrom}$, \textbf{(f)} $V_0=\SI{10}{\electronvolt\angstrom}$. }
  \label{fig:psivV}
\end{figure}

\Cref{fig:spectravV} shows the energy spectra for different values of $V_0$, and with the remaining parameters fixed at $L_{TI} = L_{MI} = \SI{50}{\nano\meter}$, $M_0^R = \SI{-0.5}{\electronvolt}$ and $h_0 = \SI{50}{\milli\electronvolt}$.
For all values of $V_0$, there are four energy bands inside the bulk gap.
The two that cross around $E = 0$ come from the topological surface states at $z = -L_{TI}$ and are therefore unaffected by the interface potential proportional to $V_0$. 
The two remaining surface states are gapped at $k_x = 0$, but this gap has an interesting, nonmonotonic dependence on $V_0$.
The interfacial surface states corresponding to $k_x = 0$ are shown in \cref{fig:psivV} for the same parameters as in \cref{fig:spectravV}.

As $V_0$ is increased, both of the energy bands corresponding to the interfacial surface states are shifted upwards in energy.
Simultaneously, the gap is also increased.
In \cref{fig:spectravV}(a), the gap between the interfacial surface states is $\SI{23}{\milli\electronvolt}$, and the gap in \cref{fig:spectravV}(b) is $\SI{31}{\milli\electronvolt}$.
This increase can be understood from the fact that the wave function is slightly shifted into the MI as $V_0$ increases.
As the interfacial surface state bands are shifted into the region bulk conduction bands, new energy bands emerge from the valence bands.

This transition can be seen in \cref{fig:spectravV}(b)-(e), where two new energy bands emerge into the bulk gap from below while the two energy bands corresponding to the initial surface states leave the gap.
As the new states enter the bulk gap from the valence band, they also become localized at the interface, as can be seen in \cref{fig:psivV}.
Moreover, the initial surface states become delocalized as $V_0$ increases, and their energies are shifted into the conduction band.
In other words, as $V_0$ increases, the surface states are shifted into the MI and are replaced by new surface states.
These new states seem to emerge from TI bulk states and they remain more localized in the TI side of the interface.
As a result, the energy gap associated with these states is much smaller.
As $V_0$ is increased further, the surface states are pushed further into the TI, reducing the energy gap even more.
This is as expected since the TI and MI should become decoupled as $V_0\to\infty$.
From \cref{fig:spectravV}(f), we can see that the energy bands corresponding to the interfacial surface states start to overlap with the energy bands associated with the surface states at $z = -L_{TI}$.

These results indicate that the interfacial barrier potential must be smaller than some cutoff in order to get an appreciable gap induced by the MI.
This can explain why it has been challenging to observe gaps in TI/MI heterostructures before the introduction of lattice-matched van der Waals structures. 
Interestingly, a small interfacial potential can be advantageous, as it increases the energy gap and shifts the surface state further into the MI.

\section{Finite size effects}
\label{sec:finite_size_effects}
\begin{figure}[htpb]
  \centering
  \includegraphics[width=0.4\textwidth]{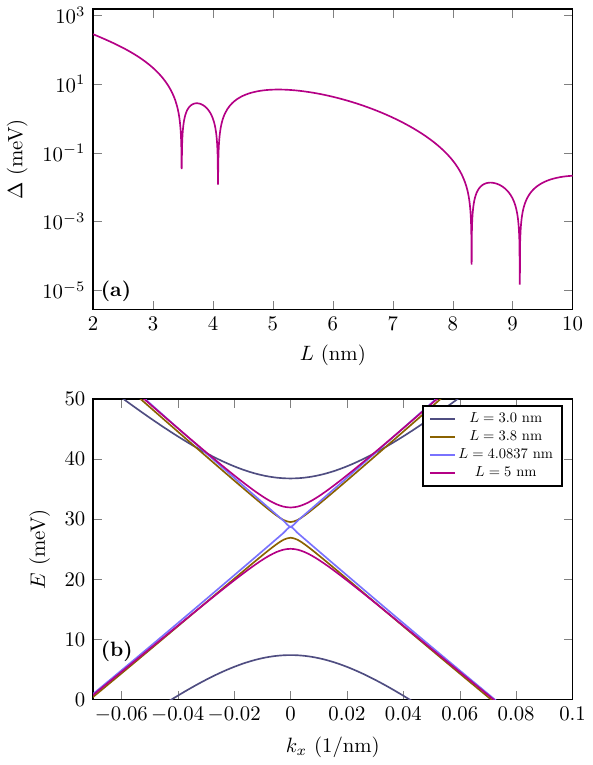}
  \caption{\textbf{(a)} Energy gap $\Delta$ i TI/MI bilayers as a function of $L = L_{TI}+L_{MI}$ with $L_{TI} = L_{MI}$, $M_0^R = \SI{-0.2}{\electronvolt}$, $h_0 = \SI{100}{\milli\electronvolt}$, and $V_0=\SI{0}{\electronvolt\angstrom}$,  \textbf{(b)} The energy spectrum for some values of $L$, illustrating that the gap closes and reopens at $L = \SI{4.0837}{\nano\meter}$.}
  \label{fig:finite_size_hetero}
\end{figure}

Next, we consider how the ferromagnetically extended surface states at $z = 0$ interact with the topological surface states at $z = -L_{TI}$ when $L_{TI}$ is small.
We let $L_{TI} = L_{MI}$ and determine the energy gap, $\Delta$, as a function of $L = L_{TI} + L_{MI}$.
The result is shown in \cref{fig:finite_size_hetero} together with the spectrum at some values of $L$.
The gap closes and reopens as a function of $L$, similar to the gap in pure \ce{Bi2Se3}~\cite{linder_prb_09, liu_prb_10, lu_prb_10}.
This is not surprising, as the TI is \ce{Bi2Se3} in our model.
However, unlike the gap in \ce{Bi2Se3}, the values of $L$ for which the gap closes are not uniformly spaced in \cref{fig:finite_size_hetero}(a).

\begin{figure}[b!]
  \centering
  \includegraphics[width=0.4\textwidth]{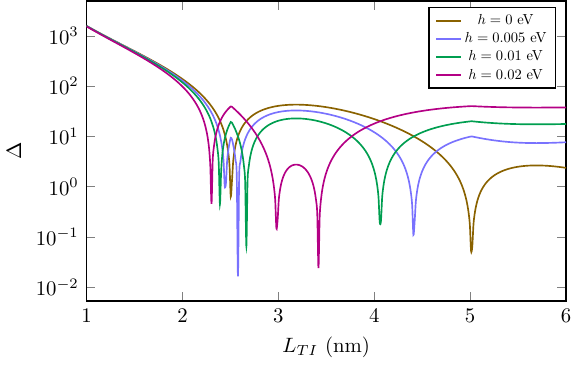}
  \caption{Energy gap in \ce{Bi2Se3} with an added constant spin splitting field $h$ as a function of $L_{TI}$.}
  \label{fig:gapsMTI}
\end{figure}

For more insight into why the ferromagnetically extended surface states give rise to nonuniformly spaced roots in the gaps, we study how an additional uniform exchange field affects the finite size effects in pure \ce{Bi2Se3}.
That is, we set $L_{MI} = 0$ and let the exchange field $h$ act also on the TI side.
The result is shown in \cref{fig:gapsMTI}.
When $h = 0$ the roots are uniformly spaced as expected.
However, as $h$ increases, the roots split, leaving a nonuniform spacing between roots.
We find that the gap as a function of $h$ in \cref{fig:gapsMTI} satisfies
\begin{equation}
  \Delta(h) = \lvert \Delta(0) - 2h \rvert.
  \label{eq:gapMTI}
\end{equation}
This can be understood by how the exchange field splits the two energy bands.
At $h = 0$, the gapped energy bands are also spin-degenerate because of the symmetries of the Hamiltonian, as explained above.
That is, there are two states with energies $E_1$ below the gap and two states with $E_2$ above the gap, such that $E_2 - E_1 = \Delta(0)$.
As $h$ is increased, each of the bands is split such that the four states have energies $E_1 \pm h$ and $E_2 \pm h$.
Hence, the gap is $\Delta(h) = E_2 - h - (E_1 + h) = \Delta(0) - 2h$.

The splitting of the roots in TI/MI heterostructures is similar, although there are some important differences.
In particular, \cref{eq:gapMTI} is not true for the TI/MI structures.
While the gap in pure magnetic TIs goes to $2h$, the gap in the TI/MI structures goes to 0 as $L\to\infty$.
Moreover, because the gap amplitude is much smaller near the second root in the pure TIs, the second root is split much more as a function of $L_{TI}$ when the exchange field is applied compared to the first root.
In contrast, the difference between the third and fourth roots in \cref{fig:finite_size_hetero} is not much larger than the difference between the first and second roots.

In both cases, it is interesting to note that, depending on the sample length, the gap can actually be reduced or even closed by increasing $h$.
This means that, although one usually expects the gap to open when reducing the temperature and going from the paramagnetic phase to the ferromagnetic phase, the opposite can happen in thin samples.

\section{Conclusion}
Summarizing, we have studied the effects of a (i) finite barrier and (ii) finite thickness on the surface states of a topological insulator/magnetic insulator (TI/MI) bilayer, as well as the effect of (ii) on an intrinsic topological magnetic insulator. We find that the energy gap of the surface states has a non-monotonic dependence on the barrier strength. Below a critical value, the barrier potential permits the surface states to extend further into the MI compared to a barrier-free interface. We also find that due to finite size effects in thin samples, an increase in the spin-splitting in the MI can diminish the gap of the surface states at certain thicknesses. This stands in contrast to the usual expectation that the gap opens as the spin-splitting increases. We hope these findings may be of use in terms of designing experiments aiming to realize extension of topological surface states into a magnetic material.

% Fakesection: Acknowledgements
\begin{acknowledgments}
This work was supported by the Research Council of Norway through Grant No. 323766 and its Centres of Excellence funding scheme Grant No. 262633 “QuSpin.” Support from Sigma2 - the National Infrastructure for High-Performance Computing and Data Storage in Norway, project NN9577K, is acknowledged.
\end{acknowledgments}

% Fakesection: Bibliogaphy
% \clearpage
\bibliography{masterref}

%merlin.mbs apsrev4-1.bst 2010-07-25 4.21a (PWD, AO, DPC) hacked
%Control: key (0)
%Control: author (72) initials jnrlst
%Control: editor formatted (1) identically to author
%Control: production of article title (-1) disabled
%Control: page (0) single
%Control: year (1) truncated
%Control: production of eprint (0) enabled
\begin{thebibliography}{21}%
\makeatletter
\providecommand \@ifxundefined [1]{%
 \@ifx{#1\undefined}
}%
\providecommand \@ifnum [1]{%
 \ifnum #1\expandafter \@firstoftwo
 \else \expandafter \@secondoftwo
 \fi
}%
\providecommand \@ifx [1]{%
 \ifx #1\expandafter \@firstoftwo
 \else \expandafter \@secondoftwo
 \fi
}%
\providecommand \natexlab [1]{#1}%
\providecommand \enquote  [1]{``#1''}%
\providecommand \bibnamefont  [1]{#1}%
\providecommand \bibfnamefont [1]{#1}%
\providecommand \citenamefont [1]{#1}%
\providecommand \href@noop [0]{\@secondoftwo}%
\providecommand \href [0]{\begingroup \@sanitize@url \@href}%
\providecommand \@href[1]{\@@startlink{#1}\@@href}%
\providecommand \@@href[1]{\endgroup#1\@@endlink}%
\providecommand \@sanitize@url [0]{\catcode `\\12\catcode `\$12\catcode
  `\&12\catcode `\#12\catcode `\^12\catcode `\_12\catcode `\%12\relax}%
\providecommand \@@startlink[1]{}%
\providecommand \@@endlink[0]{}%
\providecommand \url  [0]{\begingroup\@sanitize@url \@url }%
\providecommand \@url [1]{\endgroup\@href {#1}{\urlprefix }}%
\providecommand \urlprefix  [0]{URL }%
\providecommand \Eprint [0]{\href }%
\providecommand \doibase [0]{http://dx.doi.org/}%
\providecommand \selectlanguage [0]{\@gobble}%
\providecommand \bibinfo  [0]{\@secondoftwo}%
\providecommand \bibfield  [0]{\@secondoftwo}%
\providecommand \translation [1]{[#1]}%
\providecommand \BibitemOpen [0]{}%
\providecommand \bibitemStop [0]{}%
\providecommand \bibitemNoStop [0]{.\EOS\space}%
\providecommand \EOS [0]{\spacefactor3000\relax}%
\providecommand \BibitemShut  [1]{\csname bibitem#1\endcsname}%
\let\auto@bib@innerbib\@empty
%</preamble>
\bibitem [{\citenamefont {K\"onig}\ \emph {et~al.}(2007)\citenamefont
  {K\"onig}, \citenamefont {Wiedmann}, \citenamefont {Br\"une}, \citenamefont
  {Roth}, \citenamefont {Buhmann}, \citenamefont {Molenkamp}, \citenamefont
  {Qi},\ and\ \citenamefont {Zhang}}]{konig:07.2}%
  \BibitemOpen
  \bibfield  {author} {\bibinfo {author} {\bibfnamefont {M.}~\bibnamefont
  {K\"onig}}, \bibinfo {author} {\bibfnamefont {S.}~\bibnamefont {Wiedmann}},
  \bibinfo {author} {\bibfnamefont {C.}~\bibnamefont {Br\"une}}, \bibinfo
  {author} {\bibfnamefont {A.}~\bibnamefont {Roth}}, \bibinfo {author}
  {\bibfnamefont {H.}~\bibnamefont {Buhmann}}, \bibinfo {author} {\bibfnamefont
  {L.~W.}\ \bibnamefont {Molenkamp}}, \bibinfo {author} {\bibfnamefont
  {X.}~\bibnamefont {Qi}}, \ and\ \bibinfo {author} {\bibfnamefont
  {S.}~\bibnamefont {Zhang}},\ }\href@noop {} {\bibfield  {journal} {\bibinfo
  {journal} {Science}\ }\textbf {\bibinfo {volume} {318}},\ \bibinfo {pages}
  {766} (\bibinfo {year} {2007})}\BibitemShut {NoStop}%
\bibitem [{\citenamefont {Hasan}\ and\ \citenamefont
  {Kane}(2010)}]{Hasan:10.11}%
  \BibitemOpen
  \bibfield  {author} {\bibinfo {author} {\bibfnamefont {M.~Z.}\ \bibnamefont
  {Hasan}}\ and\ \bibinfo {author} {\bibfnamefont {C.~L.}\ \bibnamefont
  {Kane}},\ }\href@noop {} {\bibfield  {journal} {\bibinfo  {journal} {Rev.
  Mod. Phys.}\ }\textbf {\bibinfo {volume} {82}},\ \bibinfo {pages} {3045}
  (\bibinfo {year} {2010})}\BibitemShut {NoStop}%
\bibitem [{\citenamefont {Qi}\ and\ \citenamefont {Zhang}(2011)}]{Qi:11}%
  \BibitemOpen
  \bibfield  {author} {\bibinfo {author} {\bibfnamefont {X.-L.}\ \bibnamefont
  {Qi}}\ and\ \bibinfo {author} {\bibfnamefont {S.-C.}\ \bibnamefont {Zhang}},\
  }\href@noop {} {\bibfield  {journal} {\bibinfo  {journal} {Rev. Mod. Phys.}\
  }\textbf {\bibinfo {volume} {83}},\ \bibinfo {pages} {1057} (\bibinfo {year}
  {2011})}\BibitemShut {NoStop}%
\bibitem [{\citenamefont {Yu}\ \emph {et~al.}(2010)\citenamefont {Yu},
  \citenamefont {Zhang}, \citenamefont {Zhang}, \citenamefont {Zhang},
  \citenamefont {Dai},\ and\ \citenamefont {Fang}}]{yu2010}%
  \BibitemOpen
  \bibfield  {author} {\bibinfo {author} {\bibfnamefont {R.}~\bibnamefont
  {Yu}}, \bibinfo {author} {\bibfnamefont {W.}~\bibnamefont {Zhang}}, \bibinfo
  {author} {\bibfnamefont {H.-J.}\ \bibnamefont {Zhang}}, \bibinfo {author}
  {\bibfnamefont {S.-C.}\ \bibnamefont {Zhang}}, \bibinfo {author}
  {\bibfnamefont {X.}~\bibnamefont {Dai}}, \ and\ \bibinfo {author}
  {\bibfnamefont {Z.}~\bibnamefont {Fang}},\ }\href@noop {} {\bibfield
  {journal} {\bibinfo  {journal} {Science}\ }\textbf {\bibinfo {volume}
  {329}},\ \bibinfo {pages} {61} (\bibinfo {year} {2010})}\BibitemShut
  {NoStop}%
\bibitem [{\citenamefont {Nomura}\ and\ \citenamefont
  {Nagaosa}(2011)}]{Nomura:11}%
  \BibitemOpen
  \bibfield  {author} {\bibinfo {author} {\bibfnamefont {K.}~\bibnamefont
  {Nomura}}\ and\ \bibinfo {author} {\bibfnamefont {N.}~\bibnamefont
  {Nagaosa}},\ }\href@noop {} {\bibfield  {journal} {\bibinfo  {journal} {Phys.
  Rev. Lett.}\ }\textbf {\bibinfo {volume} {106}},\ \bibinfo {pages} {166802}
  (\bibinfo {year} {2011})}\BibitemShut {NoStop}%
\bibitem [{\citenamefont {Tokura}\ \emph {et~al.}(2019)\citenamefont {Tokura},
  \citenamefont {Yasuda},\ and\ \citenamefont {Tsukazaki}}]{Tokura2019}%
  \BibitemOpen
  \bibfield  {author} {\bibinfo {author} {\bibfnamefont {Y.}~\bibnamefont
  {Tokura}}, \bibinfo {author} {\bibfnamefont {K.}~\bibnamefont {Yasuda}}, \
  and\ \bibinfo {author} {\bibfnamefont {A.}~\bibnamefont {Tsukazaki}},\
  }\href@noop {} {\bibfield  {journal} {\bibinfo  {journal} {Nat. Rev. Phys.}\
  }\textbf {\bibinfo {volume} {1}},\ \bibinfo {pages} {126} (\bibinfo {year}
  {2019})}\BibitemShut {NoStop}%
\bibitem [{\citenamefont {Bernevig}\ \emph {et~al.}(2022)\citenamefont
  {Bernevig}, \citenamefont {Felser},\ and\ \citenamefont
  {Beidenkopf}}]{bernevig2022}%
  \BibitemOpen
  \bibfield  {author} {\bibinfo {author} {\bibfnamefont {B.~A.}\ \bibnamefont
  {Bernevig}}, \bibinfo {author} {\bibfnamefont {C.}~\bibnamefont {Felser}}, \
  and\ \bibinfo {author} {\bibfnamefont {H.}~\bibnamefont {Beidenkopf}},\
  }\href@noop {} {\bibfield  {journal} {\bibinfo  {journal} {Nature}\ }\textbf
  {\bibinfo {volume} {603}},\ \bibinfo {pages} {41} (\bibinfo {year}
  {2022})}\BibitemShut {NoStop}%
\bibitem [{\citenamefont {Chang}\ \emph {et~al.}(2013)\citenamefont {Chang},
  \citenamefont {Zhang}, \citenamefont {Feng}, \citenamefont {Shen},
  \citenamefont {Zhang}, \citenamefont {Guo}, \citenamefont {Li}, \citenamefont
  {Ou}, \citenamefont {Wei}, \citenamefont {Wang}, \citenamefont {Ji},
  \citenamefont {Feng}, \citenamefont {Ji}, \citenamefont {Chen}, \citenamefont
  {Jia}, \citenamefont {Dai}, \citenamefont {Fang}, \citenamefont {Zhang},
  \citenamefont {He}, \citenamefont {Wang}, \citenamefont {Lu}, \citenamefont
  {Ma},\ and\ \citenamefont {Xue}}]{Chang2013}%
  \BibitemOpen
  \bibfield  {author} {\bibinfo {author} {\bibfnamefont {C.-Z.}\ \bibnamefont
  {Chang}}, \bibinfo {author} {\bibfnamefont {J.}~\bibnamefont {Zhang}},
  \bibinfo {author} {\bibfnamefont {X.}~\bibnamefont {Feng}}, \bibinfo {author}
  {\bibfnamefont {J.}~\bibnamefont {Shen}}, \bibinfo {author} {\bibfnamefont
  {Z.}~\bibnamefont {Zhang}}, \bibinfo {author} {\bibfnamefont
  {M.}~\bibnamefont {Guo}}, \bibinfo {author} {\bibfnamefont {K.}~\bibnamefont
  {Li}}, \bibinfo {author} {\bibfnamefont {Y.}~\bibnamefont {Ou}}, \bibinfo
  {author} {\bibfnamefont {P.}~\bibnamefont {Wei}}, \bibinfo {author}
  {\bibfnamefont {L.-L.}\ \bibnamefont {Wang}}, \bibinfo {author}
  {\bibfnamefont {Z.-Q.}\ \bibnamefont {Ji}}, \bibinfo {author} {\bibfnamefont
  {Y.}~\bibnamefont {Feng}}, \bibinfo {author} {\bibfnamefont {S.}~\bibnamefont
  {Ji}}, \bibinfo {author} {\bibfnamefont {X.}~\bibnamefont {Chen}}, \bibinfo
  {author} {\bibfnamefont {J.}~\bibnamefont {Jia}}, \bibinfo {author}
  {\bibfnamefont {X.}~\bibnamefont {Dai}}, \bibinfo {author} {\bibfnamefont
  {Z.}~\bibnamefont {Fang}}, \bibinfo {author} {\bibfnamefont {S.-C.}\
  \bibnamefont {Zhang}}, \bibinfo {author} {\bibfnamefont {K.}~\bibnamefont
  {He}}, \bibinfo {author} {\bibfnamefont {Y.}~\bibnamefont {Wang}}, \bibinfo
  {author} {\bibfnamefont {L.}~\bibnamefont {Lu}}, \bibinfo {author}
  {\bibfnamefont {X.-C.}\ \bibnamefont {Ma}}, \ and\ \bibinfo {author}
  {\bibfnamefont {Q.-K.}\ \bibnamefont {Xue}},\ }\href {\doibase
  10.1126/science.1234414} {\bibfield  {journal} {\bibinfo  {journal}
  {Science}\ }\textbf {\bibinfo {volume} {340}},\ \bibinfo {pages} {167}
  (\bibinfo {year} {2013})}\BibitemShut {NoStop}%
\bibitem [{\citenamefont {Chang}\ \emph {et~al.}(2015)\citenamefont {Chang},
  \citenamefont {Zhao}, \citenamefont {Kim}, \citenamefont {Zhang},
  \citenamefont {Assaf}, \citenamefont {Heiman}, \citenamefont {Zhang},
  \citenamefont {Liu}, \citenamefont {Chan},\ and\ \citenamefont
  {Moodera}}]{Chang2015}%
  \BibitemOpen
  \bibfield  {author} {\bibinfo {author} {\bibfnamefont {C.-Z.}\ \bibnamefont
  {Chang}}, \bibinfo {author} {\bibfnamefont {W.}~\bibnamefont {Zhao}},
  \bibinfo {author} {\bibfnamefont {D.~Y.}\ \bibnamefont {Kim}}, \bibinfo
  {author} {\bibfnamefont {H.}~\bibnamefont {Zhang}}, \bibinfo {author}
  {\bibfnamefont {B.~A.}\ \bibnamefont {Assaf}}, \bibinfo {author}
  {\bibfnamefont {D.}~\bibnamefont {Heiman}}, \bibinfo {author} {\bibfnamefont
  {S.-C.}\ \bibnamefont {Zhang}}, \bibinfo {author} {\bibfnamefont
  {C.}~\bibnamefont {Liu}}, \bibinfo {author} {\bibfnamefont {M.~H.~W.}\
  \bibnamefont {Chan}}, \ and\ \bibinfo {author} {\bibfnamefont {J.~S.}\
  \bibnamefont {Moodera}},\ }\href {\doibase 10.1038/nmat4204} {\bibfield
  {journal} {\bibinfo  {journal} {Nat. Mater}\ }\textbf {\bibinfo {volume}
  {14}},\ \bibinfo {pages} {473} (\bibinfo {year} {2015})}\BibitemShut
  {NoStop}%
\bibitem [{\citenamefont {Fijalkowski}\ \emph {et~al.}(2021)\citenamefont
  {Fijalkowski}, \citenamefont {Liu}, \citenamefont {Mandal}, \citenamefont
  {Schreyeck}, \citenamefont {Brunner}, \citenamefont {Gould},\ and\
  \citenamefont {Molenkamp}}]{fijalkowski2021}%
  \BibitemOpen
  \bibfield  {author} {\bibinfo {author} {\bibfnamefont {K.~M.}\ \bibnamefont
  {Fijalkowski}}, \bibinfo {author} {\bibfnamefont {N.}~\bibnamefont {Liu}},
  \bibinfo {author} {\bibfnamefont {P.}~\bibnamefont {Mandal}}, \bibinfo
  {author} {\bibfnamefont {S.}~\bibnamefont {Schreyeck}}, \bibinfo {author}
  {\bibfnamefont {K.}~\bibnamefont {Brunner}}, \bibinfo {author} {\bibfnamefont
  {C.}~\bibnamefont {Gould}}, \ and\ \bibinfo {author} {\bibfnamefont {L.~W.}\
  \bibnamefont {Molenkamp}},\ }\href@noop {} {\bibfield  {journal} {\bibinfo
  {journal} {Nat. Comm.}\ }\textbf {\bibinfo {volume} {12}},\ \bibinfo {pages}
  {5599} (\bibinfo {year} {2021})}\BibitemShut {NoStop}%
\bibitem [{\citenamefont {Otrokov}\ \emph {et~al.}(2019)\citenamefont
  {Otrokov}, \citenamefont {Klimovskikh}, \citenamefont {Bentmann},
  \citenamefont {Estyunin}, \citenamefont {Zeugner}, \citenamefont {Aliev},
  \citenamefont {Gaß}, \citenamefont {Wolter}, \citenamefont {Koroleva},
  \citenamefont {Shikin}, \citenamefont {Blanco-Rey}, \citenamefont {Hoffmann},
  \citenamefont {Rusinov}, \citenamefont {Vyazovskaya}, \citenamefont
  {Eremeev}, \citenamefont {Koroteev}, \citenamefont {Kuznetsov}, \citenamefont
  {Freyse}, \citenamefont {Sánchez-Barriga}, \citenamefont {Amiraslanov},
  \citenamefont {Babanly}, \citenamefont {Mamedov}, \citenamefont {Abdullayev},
  \citenamefont {Zverev}, \citenamefont {Alfonsov}, \citenamefont {Kataev},
  \citenamefont {Büchner}, \citenamefont {Schwier}, \citenamefont {Kumar},
  \citenamefont {Kimura}, \citenamefont {Petaccia}, \citenamefont {Di~Santo},
  \citenamefont {Vidal}, \citenamefont {Schatz}, \citenamefont {Kißner},
  \citenamefont {Ünzelmann}, \citenamefont {Min}, \citenamefont {Moser},
  \citenamefont {Peixoto}, \citenamefont {Reinert}, \citenamefont {Ernst},
  \citenamefont {Echenique}, \citenamefont {Isaeva},\ and\ \citenamefont
  {Chulkov}}]{otrokov:19}%
  \BibitemOpen
  \bibfield  {author} {\bibinfo {author} {\bibfnamefont {M.~M.}\ \bibnamefont
  {Otrokov}}, \bibinfo {author} {\bibfnamefont {I.~I.}\ \bibnamefont
  {Klimovskikh}}, \bibinfo {author} {\bibfnamefont {H.}~\bibnamefont
  {Bentmann}}, \bibinfo {author} {\bibfnamefont {D.}~\bibnamefont {Estyunin}},
  \bibinfo {author} {\bibfnamefont {A.}~\bibnamefont {Zeugner}}, \bibinfo
  {author} {\bibfnamefont {Z.~S.}\ \bibnamefont {Aliev}}, \bibinfo {author}
  {\bibfnamefont {S.}~\bibnamefont {Gaß}}, \bibinfo {author} {\bibfnamefont
  {A.~U.~B.}\ \bibnamefont {Wolter}}, \bibinfo {author} {\bibfnamefont {A.~V.}\
  \bibnamefont {Koroleva}}, \bibinfo {author} {\bibfnamefont {A.~M.}\
  \bibnamefont {Shikin}}, \bibinfo {author} {\bibfnamefont {M.}~\bibnamefont
  {Blanco-Rey}}, \bibinfo {author} {\bibfnamefont {M.}~\bibnamefont
  {Hoffmann}}, \bibinfo {author} {\bibfnamefont {I.~P.}\ \bibnamefont
  {Rusinov}}, \bibinfo {author} {\bibfnamefont {A.~Y.}\ \bibnamefont
  {Vyazovskaya}}, \bibinfo {author} {\bibfnamefont {S.~V.}\ \bibnamefont
  {Eremeev}}, \bibinfo {author} {\bibfnamefont {Y.~M.}\ \bibnamefont
  {Koroteev}}, \bibinfo {author} {\bibfnamefont {V.~M.}\ \bibnamefont
  {Kuznetsov}}, \bibinfo {author} {\bibfnamefont {F.}~\bibnamefont {Freyse}},
  \bibinfo {author} {\bibfnamefont {J.}~\bibnamefont {Sánchez-Barriga}},
  \bibinfo {author} {\bibfnamefont {I.~R.}\ \bibnamefont {Amiraslanov}},
  \bibinfo {author} {\bibfnamefont {M.~B.}\ \bibnamefont {Babanly}}, \bibinfo
  {author} {\bibfnamefont {N.~T.}\ \bibnamefont {Mamedov}}, \bibinfo {author}
  {\bibfnamefont {N.~A.}\ \bibnamefont {Abdullayev}}, \bibinfo {author}
  {\bibfnamefont {V.~N.}\ \bibnamefont {Zverev}}, \bibinfo {author}
  {\bibfnamefont {A.}~\bibnamefont {Alfonsov}}, \bibinfo {author}
  {\bibfnamefont {V.}~\bibnamefont {Kataev}}, \bibinfo {author} {\bibfnamefont
  {B.}~\bibnamefont {Büchner}}, \bibinfo {author} {\bibfnamefont {E.~F.}\
  \bibnamefont {Schwier}}, \bibinfo {author} {\bibfnamefont {S.}~\bibnamefont
  {Kumar}}, \bibinfo {author} {\bibfnamefont {A.}~\bibnamefont {Kimura}},
  \bibinfo {author} {\bibfnamefont {L.}~\bibnamefont {Petaccia}}, \bibinfo
  {author} {\bibfnamefont {G.}~\bibnamefont {Di~Santo}}, \bibinfo {author}
  {\bibfnamefont {R.~C.}\ \bibnamefont {Vidal}}, \bibinfo {author}
  {\bibfnamefont {S.}~\bibnamefont {Schatz}}, \bibinfo {author} {\bibfnamefont
  {K.}~\bibnamefont {Kißner}}, \bibinfo {author} {\bibfnamefont
  {M.}~\bibnamefont {Ünzelmann}}, \bibinfo {author} {\bibfnamefont {C.~H.}\
  \bibnamefont {Min}}, \bibinfo {author} {\bibfnamefont {S.}~\bibnamefont
  {Moser}}, \bibinfo {author} {\bibfnamefont {T.~R.~F.}\ \bibnamefont
  {Peixoto}}, \bibinfo {author} {\bibfnamefont {F.}~\bibnamefont {Reinert}},
  \bibinfo {author} {\bibfnamefont {A.}~\bibnamefont {Ernst}}, \bibinfo
  {author} {\bibfnamefont {P.~M.}\ \bibnamefont {Echenique}}, \bibinfo {author}
  {\bibfnamefont {A.}~\bibnamefont {Isaeva}}, \ and\ \bibinfo {author}
  {\bibfnamefont {E.~V.}\ \bibnamefont {Chulkov}},\ }\href@noop {} {\bibfield
  {journal} {\bibinfo  {journal} {Nature}\ }\textbf {\bibinfo {volume} {576}},\
  \bibinfo {pages} {416} (\bibinfo {year} {2019})}\BibitemShut {NoStop}%
\bibitem [{\citenamefont {Deng}\ \emph {et~al.}(2020)\citenamefont {Deng},
  \citenamefont {Yu}, \citenamefont {Shi}, \citenamefont {Guo}, \citenamefont
  {Xu}, \citenamefont {Wang}, \citenamefont {Chen},\ and\ \citenamefont
  {Zhang}}]{deng_science_20}%
  \BibitemOpen
  \bibfield  {author} {\bibinfo {author} {\bibfnamefont {Y.}~\bibnamefont
  {Deng}}, \bibinfo {author} {\bibfnamefont {Y.}~\bibnamefont {Yu}}, \bibinfo
  {author} {\bibfnamefont {M.-Z.}\ \bibnamefont {Shi}}, \bibinfo {author}
  {\bibfnamefont {Z.}~\bibnamefont {Guo}}, \bibinfo {author} {\bibfnamefont
  {Z.}~\bibnamefont {Xu}}, \bibinfo {author} {\bibfnamefont {J.}~\bibnamefont
  {Wang}}, \bibinfo {author} {\bibfnamefont {X.}~\bibnamefont {Chen}}, \ and\
  \bibinfo {author} {\bibfnamefont {Y.}~\bibnamefont {Zhang}},\ }\href@noop {}
  {\bibfield  {journal} {\bibinfo  {journal} {Science}\ }\textbf {\bibinfo
  {volume} {367}},\ \bibinfo {pages} {895} (\bibinfo {year}
  {2020})}\BibitemShut {NoStop}%
\bibitem [{\citenamefont {Liu}\ and\ \citenamefont {Hesjedal}(2023)}]{liu2023}%
  \BibitemOpen
  \bibfield  {author} {\bibinfo {author} {\bibfnamefont {J.}~\bibnamefont
  {Liu}}\ and\ \bibinfo {author} {\bibfnamefont {T.}~\bibnamefont {Hesjedal}},\
  }\href@noop {} {\bibfield  {journal} {\bibinfo  {journal} {Advanced
  Materials}\ }\textbf {\bibinfo {volume} {35}},\ \bibinfo {pages} {2102427}
  (\bibinfo {year} {2023})}\BibitemShut {NoStop}%
\bibitem [{\citenamefont {Otrokov}\ \emph
  {et~al.}(2017{\natexlab{a}})\citenamefont {Otrokov}, \citenamefont
  {Menshchikova}, \citenamefont {Vergniory}, \citenamefont {Rusinov},
  \citenamefont {Vyazovskaya}, \citenamefont {Koroteev}, \citenamefont
  {Bihlmayer}, \citenamefont {Ernst}, \citenamefont {Echenique}, \citenamefont
  {Arnau} \emph {et~al.}}]{otrokov2017highly}%
  \BibitemOpen
  \bibfield  {author} {\bibinfo {author} {\bibfnamefont {M.~M.}\ \bibnamefont
  {Otrokov}}, \bibinfo {author} {\bibfnamefont {T.~V.}\ \bibnamefont
  {Menshchikova}}, \bibinfo {author} {\bibfnamefont {M.~G.}\ \bibnamefont
  {Vergniory}}, \bibinfo {author} {\bibfnamefont {I.~P.}\ \bibnamefont
  {Rusinov}}, \bibinfo {author} {\bibfnamefont {A.~Y.}\ \bibnamefont
  {Vyazovskaya}}, \bibinfo {author} {\bibfnamefont {Y.~M.}\ \bibnamefont
  {Koroteev}}, \bibinfo {author} {\bibfnamefont {G.}~\bibnamefont {Bihlmayer}},
  \bibinfo {author} {\bibfnamefont {A.}~\bibnamefont {Ernst}}, \bibinfo
  {author} {\bibfnamefont {P.~M.}\ \bibnamefont {Echenique}}, \bibinfo {author}
  {\bibfnamefont {A.}~\bibnamefont {Arnau}},  \emph {et~al.},\ }\href@noop {}
  {\bibfield  {journal} {\bibinfo  {journal} {2D Mater.}\ }\textbf {\bibinfo
  {volume} {4}},\ \bibinfo {pages} {025082} (\bibinfo {year}
  {2017}{\natexlab{a}})}\BibitemShut {NoStop}%
\bibitem [{\citenamefont {Otrokov}\ \emph
  {et~al.}(2017{\natexlab{b}})\citenamefont {Otrokov}, \citenamefont
  {Menshchikova}, \citenamefont {Rusinov}, \citenamefont {Vergniory},
  \citenamefont {Kuznetsov},\ and\ \citenamefont
  {Chulkov}}]{otrokov:17_magnetic}%
  \BibitemOpen
  \bibfield  {author} {\bibinfo {author} {\bibfnamefont {M.}~\bibnamefont
  {Otrokov}}, \bibinfo {author} {\bibfnamefont {T.~V.}\ \bibnamefont
  {Menshchikova}}, \bibinfo {author} {\bibfnamefont {I.}~\bibnamefont
  {Rusinov}}, \bibinfo {author} {\bibfnamefont {M.}~\bibnamefont {Vergniory}},
  \bibinfo {author} {\bibfnamefont {V.~M.}\ \bibnamefont {Kuznetsov}}, \ and\
  \bibinfo {author} {\bibfnamefont {E.~V.}\ \bibnamefont {Chulkov}},\
  }\href@noop {} {\bibfield  {journal} {\bibinfo  {journal} {JETP Lett.}\
  }\textbf {\bibinfo {volume} {105}},\ \bibinfo {pages} {297} (\bibinfo {year}
  {2017}{\natexlab{b}})}\BibitemShut {NoStop}%
\bibitem [{\citenamefont {Kagerer}\ \emph {et~al.}(2023)\citenamefont
  {Kagerer}, \citenamefont {Fornari}, \citenamefont {Buchberger}, \citenamefont
  {Tschirner}, \citenamefont {Veyrat}, \citenamefont {Kamp}, \citenamefont
  {Tcakaev}, \citenamefont {Zabolotnyy}, \citenamefont {Morelh\~ao},
  \citenamefont {Geldiyev}, \citenamefont {M\"uller}, \citenamefont {Fedorov},
  \citenamefont {Rienks}, \citenamefont {Gargiani}, \citenamefont {Valvidares},
  \citenamefont {Folkers}, \citenamefont {Isaeva}, \citenamefont {B\"uchner},
  \citenamefont {Hinkov}, \citenamefont {Claessen}, \citenamefont {Bentmann},\
  and\ \citenamefont {Reinert}}]{kagerer_prr_23}%
  \BibitemOpen
  \bibfield  {author} {\bibinfo {author} {\bibfnamefont {P.}~\bibnamefont
  {Kagerer}}, \bibinfo {author} {\bibfnamefont {C.~I.}\ \bibnamefont
  {Fornari}}, \bibinfo {author} {\bibfnamefont {S.}~\bibnamefont {Buchberger}},
  \bibinfo {author} {\bibfnamefont {T.}~\bibnamefont {Tschirner}}, \bibinfo
  {author} {\bibfnamefont {L.}~\bibnamefont {Veyrat}}, \bibinfo {author}
  {\bibfnamefont {M.}~\bibnamefont {Kamp}}, \bibinfo {author} {\bibfnamefont
  {A.~V.}\ \bibnamefont {Tcakaev}}, \bibinfo {author} {\bibfnamefont
  {V.}~\bibnamefont {Zabolotnyy}}, \bibinfo {author} {\bibfnamefont {S.~L.}\
  \bibnamefont {Morelh\~ao}}, \bibinfo {author} {\bibfnamefont
  {B.}~\bibnamefont {Geldiyev}}, \bibinfo {author} {\bibfnamefont
  {S.}~\bibnamefont {M\"uller}}, \bibinfo {author} {\bibfnamefont
  {A.}~\bibnamefont {Fedorov}}, \bibinfo {author} {\bibfnamefont
  {E.}~\bibnamefont {Rienks}}, \bibinfo {author} {\bibfnamefont
  {P.}~\bibnamefont {Gargiani}}, \bibinfo {author} {\bibfnamefont
  {M.}~\bibnamefont {Valvidares}}, \bibinfo {author} {\bibfnamefont {L.~C.}\
  \bibnamefont {Folkers}}, \bibinfo {author} {\bibfnamefont {A.}~\bibnamefont
  {Isaeva}}, \bibinfo {author} {\bibfnamefont {B.}~\bibnamefont {B\"uchner}},
  \bibinfo {author} {\bibfnamefont {V.}~\bibnamefont {Hinkov}}, \bibinfo
  {author} {\bibfnamefont {R.}~\bibnamefont {Claessen}}, \bibinfo {author}
  {\bibfnamefont {H.}~\bibnamefont {Bentmann}}, \ and\ \bibinfo {author}
  {\bibfnamefont {F.}~\bibnamefont {Reinert}},\ }\href {\doibase
  10.1103/PhysRevResearch.5.L022019} {\bibfield  {journal} {\bibinfo  {journal}
  {Phys. Rev. Res.}\ }\textbf {\bibinfo {volume} {5}},\ \bibinfo {pages}
  {L022019} (\bibinfo {year} {2023})}\BibitemShut {NoStop}%
\bibitem [{\citenamefont {Zhang}\ \emph {et~al.}(2009)\citenamefont {Zhang},
  \citenamefont {Liu}, \citenamefont {Qi}, \citenamefont {Dai}, \citenamefont
  {Fang},\ and\ \citenamefont {Zhang}}]{zhang2009}%
  \BibitemOpen
  \bibfield  {author} {\bibinfo {author} {\bibfnamefont {H.}~\bibnamefont
  {Zhang}}, \bibinfo {author} {\bibfnamefont {C.-X.}\ \bibnamefont {Liu}},
  \bibinfo {author} {\bibfnamefont {X.-L.}\ \bibnamefont {Qi}}, \bibinfo
  {author} {\bibfnamefont {X.}~\bibnamefont {Dai}}, \bibinfo {author}
  {\bibfnamefont {Z.}~\bibnamefont {Fang}}, \ and\ \bibinfo {author}
  {\bibfnamefont {S.-C.}\ \bibnamefont {Zhang}},\ }\href {\doibase
  10.1038/nphys1270} {\bibfield  {journal} {\bibinfo  {journal} {Nat. Phys.}\
  }\textbf {\bibinfo {volume} {5}},\ \bibinfo {pages} {438} (\bibinfo {year}
  {2009})}\BibitemShut {NoStop}%
\bibitem [{\citenamefont {Zhang}\ \emph {et~al.}(2019)\citenamefont {Zhang},
  \citenamefont {Shi}, \citenamefont {Zhu}, \citenamefont {Xing}, \citenamefont
  {Zhang},\ and\ \citenamefont {Wang}}]{zhang_prl_19}%
  \BibitemOpen
  \bibfield  {author} {\bibinfo {author} {\bibfnamefont {D.}~\bibnamefont
  {Zhang}}, \bibinfo {author} {\bibfnamefont {M.}~\bibnamefont {Shi}}, \bibinfo
  {author} {\bibfnamefont {T.}~\bibnamefont {Zhu}}, \bibinfo {author}
  {\bibfnamefont {D.}~\bibnamefont {Xing}}, \bibinfo {author} {\bibfnamefont
  {H.}~\bibnamefont {Zhang}}, \ and\ \bibinfo {author} {\bibfnamefont
  {J.}~\bibnamefont {Wang}},\ }\href {\doibase 10.1103/PhysRevLett.122.206401}
  {\bibfield  {journal} {\bibinfo  {journal} {Phys. Rev. Lett.}\ }\textbf
  {\bibinfo {volume} {122}},\ \bibinfo {pages} {206401} (\bibinfo {year}
  {2019})}\BibitemShut {NoStop}%
\bibitem [{\citenamefont {Linder}\ \emph {et~al.}(2009)\citenamefont {Linder},
  \citenamefont {Yokoyama},\ and\ \citenamefont {Sudb\o{}}}]{linder_prb_09}%
  \BibitemOpen
  \bibfield  {author} {\bibinfo {author} {\bibfnamefont {J.}~\bibnamefont
  {Linder}}, \bibinfo {author} {\bibfnamefont {T.}~\bibnamefont {Yokoyama}}, \
  and\ \bibinfo {author} {\bibfnamefont {A.}~\bibnamefont {Sudb\o{}}},\ }\href
  {\doibase 10.1103/PhysRevB.80.205401} {\bibfield  {journal} {\bibinfo
  {journal} {Phys. Rev. B}\ }\textbf {\bibinfo {volume} {80}},\ \bibinfo
  {pages} {205401} (\bibinfo {year} {2009})}\BibitemShut {NoStop}%
\bibitem [{\citenamefont {Liu}\ \emph {et~al.}(2010)\citenamefont {Liu},
  \citenamefont {Zhang}, \citenamefont {Yan}, \citenamefont {Qi}, \citenamefont
  {Frauenheim}, \citenamefont {Dai}, \citenamefont {Fang},\ and\ \citenamefont
  {Zhang}}]{liu_prb_10}%
  \BibitemOpen
  \bibfield  {author} {\bibinfo {author} {\bibfnamefont {C.-X.}\ \bibnamefont
  {Liu}}, \bibinfo {author} {\bibfnamefont {H.}~\bibnamefont {Zhang}}, \bibinfo
  {author} {\bibfnamefont {B.}~\bibnamefont {Yan}}, \bibinfo {author}
  {\bibfnamefont {X.-L.}\ \bibnamefont {Qi}}, \bibinfo {author} {\bibfnamefont
  {T.}~\bibnamefont {Frauenheim}}, \bibinfo {author} {\bibfnamefont
  {X.}~\bibnamefont {Dai}}, \bibinfo {author} {\bibfnamefont {Z.}~\bibnamefont
  {Fang}}, \ and\ \bibinfo {author} {\bibfnamefont {S.-C.}\ \bibnamefont
  {Zhang}},\ }\href {\doibase 10.1103/PhysRevB.81.041307} {\bibfield  {journal}
  {\bibinfo  {journal} {Phys. Rev. B}\ }\textbf {\bibinfo {volume} {81}},\
  \bibinfo {pages} {041307} (\bibinfo {year} {2010})}\BibitemShut {NoStop}%
\bibitem [{\citenamefont {Lu}\ \emph {et~al.}(2010)\citenamefont {Lu},
  \citenamefont {Shan}, \citenamefont {Yao}, \citenamefont {Niu},\ and\
  \citenamefont {Shen}}]{lu_prb_10}%
  \BibitemOpen
  \bibfield  {author} {\bibinfo {author} {\bibfnamefont {H.-Z.}\ \bibnamefont
  {Lu}}, \bibinfo {author} {\bibfnamefont {W.-Y.}\ \bibnamefont {Shan}},
  \bibinfo {author} {\bibfnamefont {W.}~\bibnamefont {Yao}}, \bibinfo {author}
  {\bibfnamefont {Q.}~\bibnamefont {Niu}}, \ and\ \bibinfo {author}
  {\bibfnamefont {S.-Q.}\ \bibnamefont {Shen}},\ }\href {\doibase
  10.1103/PhysRevB.81.115407} {\bibfield  {journal} {\bibinfo  {journal} {Phys.
  Rev. B}\ }\textbf {\bibinfo {volume} {81}},\ \bibinfo {pages} {115407}
  (\bibinfo {year} {2010})}\BibitemShut {NoStop}%
\end{thebibliography}%

\end{document}